# The dual frequency RV-coupling coefficient: a novel measure for quantifying cross-frequency information transactions in the brain

RD Pascual-Marqui[1,2], P Faber[1], T Kinoshita[2], Y Kitaura[2], K Kochi[1], P Milz[1], K Nishida[2], M Yoshimura[2]

[1]The KEY Institute for Brain-Mind Research, University of Zurich, Switzerland; [2]Department of Neuropsychiatry, Kansai Medical University, Osaka, Japan

**Corresponding author:** RD Pascual-Marqui
www.uzh.ch/keyinst/loreta.htm
scholar.google.com/citations?user=pascualmarqui
www.ncbi.nlm.nih.gov/pubmed?term=pascual-marqui[Author]

## 1. Abstract

Identifying dynamic transactions between brain regions has become increasingly important. Measurements within and across brain structures, demonstrating the occurrence of bursts of beta/gamma oscillations only during one specific phase of each theta/alpha cycle, have motivated the need to advance beyond linear and stationary time series models. Here we offer a novel measure, namely, the "dual frequency RV-coupling coefficient", for assessing different types of frequency-frequency interactions that subserve information flow in the brain. This is a measure of coherence between two complex-valued vectors, consisting of the set of Fourier coefficients for two different frequency bands, within or across two brain regions. RV-coupling is expressed in terms of instantaneous and lagged components. Furthermore, by using normalized Fourier coefficients (unit modulus), phase-type couplings can also be measured. The dual frequency RV-coupling coefficient is based on previous work: the second order bispectrum, i.e. the dual-frequency coherence (Thomson 1982; Haykin & Thomson 1998); the RV-coefficient (Escoufier 1973); Gorrostieta et al (2012); and Pascual-Marqui et al (2011). This paper presents the new measure, and outlines relevant statistical tests. The novel aspects of the "dual frequency RV-coupling coefficient" are: (1) it can be applied to two multivariate time series; (2) the method is not limited to single discrete frequencies, and in addition, the frequency bands are treated by means of appropriate multivariate statistical methodology; (3) the method makes use of a novel generalization of the RV-coefficient for complex-valued multivariate data; (4) real and imaginary covariance contributions to the RV-coherence are obtained, allowing the definition of a "lagged-coupling" measure that is minimally affected by the low spatial resolution of estimated cortical electric neuronal activity.

## 2. Introduction

Informally, it is said that two brain regions are "connected" if their activity time series are "similar" (Worsley et al., 2005). A very simple hypothetical example consists of using the coherence as a measure of "similarity", applied to short-range, local bipolar ECoG (electrocorticogram) recordings from two distant cortical sites. The coherence is certainly very useful, but it may not be sufficient for uncovering connection mechanisms that fall outside the realm of linear and stationary processes.

An excellent illustration of "associations", i.e. of "connections", that can occur under non-stationarity and non-linearity can be found in Figure 1 in (Jirsa and Muller, 2013). For instance, the signals in Figure 1A and 1D in (Jirsa and Muller, 2013) exhibit very strong phase to envelope (i.e. instantaneous amplitude) correlation, which is a form of "similarity" (i.e. connectivity) not captured by the coherence between the raw signals. The existence of this type of "connection", known as "phase-amplitude cross-frequency coupling", has been experimentally demonstrated in animal studies, see e.g. (Chrobak and Buzsaki, 1998;Buzsaki et al., 2003), and in humans, see e.g. (Canolty et al., 2006;Lega et al., 2014). Furthermore, in the case of theta/gamma phase-phase coupling, it has been demonstrated that it takes place not only within a structure (i.e. within a single signal), but also across structures (Belluscio et al., 2012).

A review of published methods currently being used for quantifying "connectivity" under these particular forms of non-linear and non-stationary conditions can be found in (Penny et al., 2008;van Wijk et al., 2015). The three most commonly studied forms of cross-frequency coupling are:
1. Amplitude-amplitude, see e.g. (Bruns et al., 2000;Bruns and Eckhorn, 2004).
2. Phase-amplitude, see e.g. (Canolty et al., 2006;Penny et al., 2008;Ozkurt and Schnitzler, 2011).
3. Phase-phase, see e.g. (Lachaux et al., 1999;Chaieb et al., 2015).

Almost all the methods quoted above make use of the analytic signal (obtained with the Hilbert transform), which provides time varying signals of instantaneous amplitude (i.e. the envelope) and instantaneous phase, which are then used for measuring coupling.

Surprisingly, however, what would appear to be the simplest method for quantifying cross-frequency associations is hardly used at all, namely, the "second order bispectrum" defined by (Thomson, 1982), on Page 1089, Equation 4 therein. This measure is now more commonly known as the "dual-frequency coherence" (Haykin and Thomson, 1998;Mellors et al., 1998). It consists of calculating the coherence between the discrete Fourier coefficients at two frequencies. When the frequencies are the same, this gives the common coherence. However, when the frequencies are different, it gives a simple and straightforward general measure of cross-frequency coupling, which includes both amplitude and phase information.

The type of non-stationary processes that are particularly well characterized by the dual frequency coherence correspond to harmonizable or cyclostationary processes (Lii and Rosenblatt, 2002;Olhede and Ombao, 2013). In other more general cases, the dual frequency coherence still provides important, albeit possibly not complete, information.

The complex-valued covariance between Fourier coefficients at different frequencies, upon which the dual-frequency coherence is based, has recently been proposed as a statistic for testing second order stationarity in multivariate time series (Jentsch and Rao, 2015). This result constitutes a strong theoretical foundation for the use of the dual-frequency coherence as a general measure of non-stationary cross-frequency coupling.

A recent application using the dual-frequency coherence in the analysis of brain electric signals can be found in (Gorrostieta et al., 2012).

It is the aim of this work to offer a novel generalized version of the dual-frequency coherence for measuring dynamic transactions of information between brain regions.

### 3. The complex-valued cross-covariance between Fourier coefficients at two frequencies (second order bispectrum) and its normalized form (dual-frequency coherence)

Let $x_k(t)$ and $y_k(t)$ denote two univariate time series, for discrete time $t = 0...(N_T - 1)$, with $k = 1...N_R$ denoting the k-th time segment (i.e. epoch or time window). Let:

**Eq. 1** $$x_k(\omega) = \sum_{t=0}^{N_T-1} x_k(t) e^{-\iota 2\pi \omega t/N_T}$$

and:

**Eq. 2** $$y_k(\omega) = \sum_{t=0}^{N_T-1} y_k(t) e^{-\iota 2\pi \omega t/N_T}$$

denote the discrete Fourier transform (DFT) of "x" and "y" respectively, for discrete frequency $\omega = 0 ... N_T/2$, with $\iota = \sqrt{-1}$.

Let (see e.g. (Thomson, 1982), Page 1089, Equation 4 therein):

**Eq. 3** $$s_{xy}(\omega_1, \omega_2) = \frac{1}{N_R} \sum_{k=1}^{N_R} x_k(\omega_1) y_k^*(\omega_2)$$

denote the covariance between the DFT of "x" at frequency $\omega_1$, with the DFT of "y" at frequency $\omega_2$. In general, the superscript "*" denotes conjugate-transpose.

Note that $s_{xx}(\omega_1, \omega_1)$ corresponds to the periodogram estimator for the spectral density of "x" at frequency $\omega_1$. Also note that if "x" is second order stationary and $\omega_1 \neq \omega_2$, then asymptotically $s_{xx}(\omega_1, \omega_2) = 0$, see e.g. (Jentsch and Rao, 2015). Similar results hold for "y". Also note that $s_{xy}(\omega_1, \omega_1)$ is the classical Hermitian complex-valued covariance between "x" and "y" at frequency $\omega_1$.

In general, $s_{xy}(\omega_1, \omega_2)$ in Eq. 3 corresponds to the "second order bispectrum" of (Thomson, 1982). In its normalized form, it is the "dual-frequency coherence" (Haykin and Thomson, 1998;Mellors et al., 1998):

**Eq. 4** $$c_{xy}^2(\omega_1, \omega_2) = \frac{\{\text{Re}[s_{xy}(\omega_1, \omega_2)]\}^2 + \{\text{Im}[s_{xy}(\omega_1, \omega_2)]\}^2}{s_{xx}(\omega_1, \omega_1) s_{yy}(\omega_2, \omega_2)} = \frac{|s_{xy}(\omega_1, \omega_2)|^2}{s_{xx}(\omega_1, \omega_1) s_{yy}(\omega_2, \omega_2)}$$

The actual definitions of the "second order bispectrum" (Thomson, 1982) and dual-frequency coherence (Haykin and Thomson, 1998;Mellors et al., 1998) make heavy use of a multitaper method. In this work, in Eq. 3 and Eq. 4, we use the simple periodogram estimators.

A recent application of the dual-frequency coherence in the analysis of brain electric signals can be found in (Gorrostieta et al., 2012).

## 4. The RV-coefficient

The RV-coefficient was proposed by Escoufier (Escoufier, 1973;Robert and Escoufier, 1976) as a measure of association between two real valued random vectors. It is a generalization of the bivariate correlation coefficient. A recent review of measures of association, which includes an in-depth evaluation of the RV coefficient, can be found in (Josse and Holmes, 2013).

Consider the case of two real-valued multivariate random vectors with zero mean, $\mathbf{U}_i \in \mathbb{R}^{p \times 1}$ and $\mathbf{V}_i \in \mathbb{R}^{q \times 1}$, corresponding to samples $i = 1 ... N_S$. In this context, the RV-coefficient is defined as:

**Eq. 5** $$RV_{uv} = \frac{tr(\mathbf{S}_{uv} \mathbf{S}_{uv}^T)}{\sqrt{tr(\mathbf{S}_{uu}^2) tr(\mathbf{S}_{vv}^2)}}$$

where $tr(\mathbf{M})$ denotes the trace of the matrix $\mathbf{M}$, the superscript "$T$" denotes matrix transpose, and where:

Eq. 6 $\quad \mathbf{S}_{uu} = \frac{1}{N_S} \sum_{i=1}^{N_S} \mathbf{U}_i \mathbf{U}_i^T \; ; \; \mathbf{S}_{vv} = \frac{1}{N_S} \sum_{i=1}^{N_S} \mathbf{V}_i \mathbf{V}_i^T \; ; \; \mathbf{S}_{uv} = \frac{1}{N_S} \sum_{i=1}^{N_S} \mathbf{U}_i \mathbf{V}_i^T$

denote the usual estimated covariance matrices for zero mean data.

The RV-coefficient is equivalent to the squared correlation coefficient for the case when $p=q=1$. In general, it takes values between zero and one.

## 5. The complex-valued RV-coefficient

In the case of centered (i.e. zero mean), complex-valued data, with $\mathbf{U}_i \in \mathbb{C}^{p \times 1}$ and $\mathbf{V}_i \in \mathbb{C}^{q \times 1}$, we will use the following straightforward generalization and definition for the RV-coefficient (which to the best of our knowledge hasn't been previously published):

Eq. 7 $\quad RV_{uv} = \frac{tr(\mathbf{S}_{uv} \mathbf{S}_{uv}^*)}{\sqrt{tr(\mathbf{S}_{uu}^2) tr(\mathbf{S}_{vv}^2)}}$

with:

Eq. 8 $\quad \mathbf{S}_{uu} = \frac{1}{N_S} \sum_{i=1}^{N_S} \mathbf{U}_i \mathbf{U}_i^* \; ; \; \mathbf{S}_{vv} = \frac{1}{N_S} \sum_{i=1}^{N_S} \mathbf{V}_i \mathbf{V}_i^* \; ; \; \mathbf{S}_{uv} = \frac{1}{N_S} \sum_{i=1}^{N_S} \mathbf{U}_i \mathbf{V}_i^*$

In Eq. 8, the covariance matrices are complex-valued. In particular, $\mathbf{S}_{uu}$ and $\mathbf{S}_{vv}$ are Hermitian non-negative definite matrices, satisfying the condition $\mathbf{M} = \mathbf{M}^*$ where (as previously stated) the superscript "*" denotes conjugate-transpose. Furthermore, the matrix $(\mathbf{S}_{uv} \mathbf{S}_{uv}^*)$ in Eq. 7 is also Hermitian non-negative definite.

Note that the RV-coefficient for complex-valued data is equivalent to the squared modulus of the complex valued coherence for the case when $p=q=1$. In general, it takes values between zero and one.

## 6. The "dual frequency RV-coupling coefficient": dual-frequency coherence in the case of broad bands

The definition of the RV-coefficient for complex valued data (Eq. 7 and Eq. 8) can now be used for computing a generalized measure of dual-frequency coherence, for the case of broad bands, as follows.

Let "x" and "y" denote two univariate time series. Let $\boldsymbol{\theta} = (\theta_1, \theta_2, ..., \theta_p)$ denote the set of "p" discrete frequencies that define the frequency band of interest for "x"; and let $\boldsymbol{\beta} = (\beta_1, \beta_2, ..., \beta_q)$ denote the set of "q" discrete frequencies that define the frequency band of interest for "y".

Based on Eq. 1 and Eq. 2, define the complex valued vectors formed by the DFT coefficients:

Eq. 9  $\mathbf{X}_k(\boldsymbol{\theta}) = \begin{pmatrix} x_k(\theta_1) \\ x_k(\theta_2) \\ ... \\ x_k(\theta_p) \end{pmatrix} \in \mathbb{C}^{p \times 1}$

and:

Eq. 10  $\mathbf{Y}_k(\boldsymbol{\beta}) = \begin{pmatrix} y_k(\beta_1) \\ y_k(\beta_2) \\ ... \\ y_k(\beta_q) \end{pmatrix} \in \mathbb{C}^{q \times 1}$

where $k = 1...N_R$ denotes the k-th time segment (i.e. epoch or time window). Plugging Eq. 9 and Eq. 10 into Eq. 7 and Eq. 8 gives the "dual frequency RV-coupling coefficient":

Eq. 11  $RV_{xy}(\boldsymbol{\theta},\boldsymbol{\beta}) = \dfrac{tr\left[\mathbf{S}_{xy}(\boldsymbol{\theta},\boldsymbol{\beta})\mathbf{S}_{xy}^*(\boldsymbol{\theta},\boldsymbol{\beta})\right]}{\sqrt{tr\left[\mathbf{S}_{xx}^2(\boldsymbol{\theta},\boldsymbol{\theta})\right]tr\left[\mathbf{S}_{yy}^2(\boldsymbol{\beta},\boldsymbol{\beta})\right]}}$

Eq. 12  $\mathbf{S}_{xx}(\boldsymbol{\theta},\boldsymbol{\theta}) = \dfrac{1}{N_R}\sum_{k=1}^{N_R} \mathbf{X}_k(\boldsymbol{\theta})\mathbf{X}_k^*(\boldsymbol{\theta})$

Eq. 13  $\mathbf{S}_{yy}(\boldsymbol{\beta},\boldsymbol{\beta}) = \dfrac{1}{N_R}\sum_{k=1}^{N_R} \mathbf{Y}_k(\boldsymbol{\beta})\mathbf{Y}_k^*(\boldsymbol{\beta})$

Eq. 14  $\mathbf{S}_{xy}(\boldsymbol{\theta},\boldsymbol{\beta}) = \dfrac{1}{N_R}\sum_{k=1}^{N_R} \mathbf{X}_k(\boldsymbol{\theta})\mathbf{Y}_k^*(\boldsymbol{\beta})$

Note that the "dual frequency RV-coupling coefficient" in Eq. 11 is equivalent to the "dual-frequency coherence" in Eq. 4 for the case of single discrete frequencies, i.e. when $p = q = 1$.

It is also important to note that the "frequency band" approach used here (based on Eq. 9, Eq. 10, and Eq. 11) is very different from the "frequency band" approach used in the study by (Gorrostieta et al., 2012).

In our multivariate approach, the DFT coefficients for the discrete frequencies that constitute each frequency band are treated as the set of variables that form a random vector, which implies that all variances and covariances between all pairs of discrete frequencies are accounted for. On the other hand, (Gorrostieta et al., 2012) use the classic dual-frequency coherence formula applied to the average DFT over the discrete frequencies that constitute each frequency band.

### 7. Components of the "dual frequency RV-coupling coefficient": real and imaginary parts, instantaneous and lagged connectivity

Note that the numerator in Eq. 11 can be written as:

Eq. 15  $tr\left[\mathbf{S}_{xy}(\boldsymbol{\theta},\boldsymbol{\beta})\mathbf{S}_{xy}^*(\boldsymbol{\theta},\boldsymbol{\beta})\right] = \sum_{i=1}^{p}\sum_{j=1}^{q}\left|\left[\mathbf{S}_{xy}(\boldsymbol{\theta},\boldsymbol{\beta})\right]_{ij}\right|^2 = \sum_{i=1}^{p}\sum_{j=1}^{q}\left\{\mathrm{Re}\left[\mathbf{S}_{xy}(\boldsymbol{\theta},\boldsymbol{\beta})\right]_{ij}\right\}^2 + \sum_{i=1}^{p}\sum_{j=1}^{q}\left\{\mathrm{Im}\left[\mathbf{S}_{xy}(\boldsymbol{\theta},\boldsymbol{\beta})\right]_{ij}\right\}^2$

where $\left[\mathbf{S}_{xy}(\boldsymbol{\theta},\boldsymbol{\beta})\right]_{ij}$ denotes the complex-valued element (i,j) of the matrix $\mathbf{S}_{xy}(\boldsymbol{\theta},\boldsymbol{\beta})$, with $i = 1...p$, and $j = 1...q$, and $\mathrm{Re}[\bullet]$ and $\mathrm{Im}[\bullet]$ denote the real and imaginary parts of the argument.

This gives a natural decomposition of the "dual frequency RV-coupling coefficient" into contributions from the real and imaginary parts of the covariance. Thus, Eq. 11 can meaningfully be written as:

**Eq. 16** $\quad RV_{xy}(\theta,\beta) = RV_{xy}^{Re}(\theta,\beta) + RV_{xy}^{Im}(\theta,\beta)$

with the real and imaginary contributions defined as:

**Eq. 17** $\quad RV_{xy}^{Re}(\theta,\beta) = \dfrac{\sum_{i=1}^{p}\sum_{j=1}^{q}\left\{Re\left[\mathbf{S}_{xy}(\theta,\beta)\right]_{ij}\right\}^2}{\sqrt{tr\left[\mathbf{S}_{xx}^2(\theta,\theta)\right]tr\left[\mathbf{S}_{yy}^2(\beta,\beta)\right]}}$

**Eq. 18** $\quad RV_{xy}^{Im}(\theta,\beta) = \dfrac{\sum_{i=1}^{p}\sum_{j=1}^{q}\left\{Im\left[\mathbf{S}_{xy}(\theta,\beta)\right]_{ij}\right\}^2}{\sqrt{tr\left[\mathbf{S}_{xx}^2(\theta,\theta)\right]tr\left[\mathbf{S}_{yy}^2(\beta,\beta)\right]}}$

These components (Eq. 17 and Eq. 18) are equivalent, respectively, to the squares of the real and imaginary parts of the coherence for the case $p=q=1$ (see Eq. 4).

It is well known that signals of electric neuronal activity estimated from extracranial EEG / MEG recordings have low spatial resolution, see e.g. (Pascual-Marqui, 2007;Pascual-Marqui et al., 2011). This implies that the signals will be highly correlated at lag zero (i.e. instantaneously). It is of interest to take this into account, and to develop measures of connectivity that reflect physiology, without being confounded with this low resolution artifact.

One solution to this problem, as proposed by (Nolte et al., 2004), is to consider only the imaginary part of the coherence. In analogy with this approach, the imaginary component of the "dual frequency RV-coupling coefficient" (Eq. 18) can be used.

Another solution to this problem, as proposed by (Pascual-Marqui, 2007;Pascual-Marqui et al., 2011), consists of expressing the total "connectivity" in terms of an instantaneous component and a non-instantaneous (i.e. lagged) component. The instantaneous component, which contains the low spatial resolution artifact, corresponds to the real component of the "dual frequency RV-coupling coefficient" (Eq. 17). The physiological lagged (non-instantaneous) connectivity is:

**Eq. 19** $\quad RV_{xy}^{Lag}(\theta,\beta) = \dfrac{RV_{xy}^{Im}(\theta,\beta)}{1 - RV_{xy}^{Re}(\theta,\beta)}$

in analogy with Equation 3.17 in (Pascual-Marqui et al., 2011).

## 8. The "dual frequency RV-coupling coefficient" between multivariate time series

A further generalization of the "dual frequency RV-coupling coefficient" corresponds to the case when the univariate time series "x" and "y" in Eq. 1 and Eq. 2, and in Eq. 9 and Eq. 10 are in fact multivariate time series. This situation may arise, for instance, when considering estimated signals of cortical electric neuronal activity from EEG recordings. Most inverse EEG solutions produce a 3-dimensional time series at each cortical location, consisting of the three components of the current density vector field produced by the electric neuronal activity, see e.g. (Pascual-Marqui, 2009).

Consider the general multivariate case, with multivariate time series $\mathbf{X}_k(t) \in \mathbb{R}^{r \times 1}$ and $\mathbf{Y}_k(t) \in \mathbb{R}^{s \times 1}$, and with corresponding Fourier transforms $\mathbf{X}_k(\omega) \in \mathbb{C}^{r \times 1}$ and $\mathbf{Y}_k(\omega) \in \mathbb{C}^{s \times 1}$. Then the complex-valued vectors for frequency bands in Eq. 9 now consist of "p" stacked vectors of dimension

"r" (corresponding to the "r" components of "**X**" at the "p" discrete frequencies that compose the θ band), which is a vector with $(pr)$ components:

Eq. 20 $\quad \mathbf{W}_k(\theta) = \begin{pmatrix} \mathbf{X}_k(\theta_1) \\ \mathbf{X}_k(\theta_2) \\ \ldots \\ \mathbf{X}_k(\theta_p) \end{pmatrix} \in \mathbb{C}^{(pr) \times 1}$

Similarly, the complex-valued vectors for frequency bands in Eq. 10 consist of "q" stacked vectors of dimension "s" (corresponding to the "s" components of "**Y**" at "q" discrete frequencies that compose the β band), which is a vector with $(qs)$ components:

Eq. 21 $\quad \mathbf{Z}_k(\beta) = \begin{pmatrix} \mathbf{Y}_k(\beta_1) \\ \mathbf{Y}_k(\beta_2) \\ \ldots \\ \mathbf{Y}_k(\beta_q) \end{pmatrix} \in \mathbb{C}^{(qs) \times 1}$

Plugging Eq. 20 and Eq. 21 into Eq. 11 gives:

Eq. 22 $\quad RV_{wz}(\theta, \beta) = \dfrac{tr\left[\mathbf{S}_{wz}(\theta, \beta) \mathbf{S}_{wz}^{*}(\theta, \beta)\right]}{\sqrt{tr\left[\mathbf{S}_{ww}^{2}(\theta, \theta)\right] tr\left[\mathbf{S}_{zz}^{2}(\beta, \beta)\right]}}$

which is the general dual frequency RV-coupling coefficient between two multivariate time series at two frequency bands.

## 9. Phase-phase coupling, and phase-amplitude-phase coupling, and generalized coupling

The dual frequency RV-coupling coefficient as defined in Eq. 9, Eq. 10, and Eq. 11 corresponds to generalized coupling, which takes into account both the amplitude and the phase information implicit in the DFT coefficients. Therefore, this measure should be sensitive, but not specific, to any form of coupling: amplitude-amplitude, phase-phase, and phase-amplitude.

Now consider the case where the Fourier coefficients for the time series "x" in Eq. 9 are normalized to unit modulus. This can be achieved by changing Eq. 9 to:

Eq. 23 $\quad \breve{\mathbf{X}}_k(\theta) = \begin{pmatrix} x_k(\theta_1)/|x_k(\theta_1)| \\ x_k(\theta_2)/|x_k(\theta_2)| \\ \ldots \\ x_k(\theta_p)/|x_k(\theta_p)| \end{pmatrix} \in \mathbb{C}^{p \times 1}$

Similarly, transforming Eq. 10, we have:

Eq. 24 $\quad \breve{\mathbf{Y}}_k(\theta) = \begin{pmatrix} y_k(\beta_1)/|y_k(\beta_1)| \\ y_k(\beta_2)/|y_k(\beta_2)| \\ \ldots \\ y_k(\beta_q)/|y_k(\beta_q)| \end{pmatrix} \in \mathbb{C}^{q \times 1}$

Eq. 23 and Eq. 24 contain only phase information for both time series "x" and "y". Plugging these phase-only data into Eq. 12, Eq. 13, Eq. 14, and Eq. 11 gives the "phase-phase dual frequency RV-coupling coefficient":

Eq. 25 $$RV_{\bar{x}\bar{y}}(\theta,\beta) = \frac{tr\left[\mathbf{S}_{\bar{x}\bar{y}}(\theta,\beta)\mathbf{S}^{*}_{\bar{x}\bar{y}}(\theta,\beta)\right]}{\sqrt{tr\left[\mathbf{S}^{2}_{\bar{x}\bar{x}}(\theta,\theta)\right]tr\left[\mathbf{S}^{2}_{\bar{y}\bar{y}}(\beta,\beta)\right]}}$$

This measure (Eq. 25) is specifically tailored to detect phase-phase coupling between frequency bands. Note that this measure can be expressed in terms of instantaneous and lagged components, as above (Eq. 17, Eq. 18, and Eq. 19). Also note that the "phase-phase dual frequency RV-coupling coefficient" is equivalent to the classic phase locking value (Lachaux et al., 1999; Penny et al., 2008) for the case when $p = q = 1$.

Now consider the mixture of phase-only information for time series "x" (Eq. 23), with full amplitude-phase information for time series "y" (Eq. 10), plugged into Eq. 12, Eq. 13, Eq. 14, and Eq. 11. This gives the "phase-amplitude-phase dual frequency RV-coupling coefficient":

Eq. 26 $$RV_{\bar{x}y}(\theta,\beta) = \frac{tr\left[\mathbf{S}_{\bar{x}y}(\theta,\beta)\mathbf{S}^{*}_{\bar{x}y}(\theta,\beta)\right]}{\sqrt{tr\left[\mathbf{S}^{2}_{\bar{x}\bar{x}}(\theta,\theta)\right]tr\left[\mathbf{S}^{2}_{yy}(\beta,\beta)\right]}}$$

Note that this definition (Eq. 10, Eq. 23, Eq. 26) for phase-amplitude-phase coupling does not make use of pure amplitude information for the "y" time series, as is commonly used in the phase-amplitude coupling literature, see e.g. (Penny et al., 2008). Nevertheless, this new measure proposed here will also detect phase-amplitude coupling, albeit in a different way. In addition, note that, as before, this measure can be expressed in terms of instantaneous and lagged components (Eq. 17, Eq. 18, and Eq. 19).

In general, it is essential to exercise caution when using and interpreting these different forms of coupling, because as has been clearly pointed out by (Hyafil, 2015), they can be confounded with one another (for instance, phase-amplitude coupling might be due to phase-frequency coupling).

## 10. Dynamic, time-varying, dual frequency RV-coupling coefficient

Since the dual frequency RV-coupling coefficient is essentially based on the discrete Fourier transform, it can be adapted to track time varying coupling by using a sliding short time Fourier transform (STFT). This is one of the simplest ways to track dynamic changes, which is a commonly used method in the analysis of time-varying spectra, see e.g. (Cohen, 1995).

## 11. Statistics

Consider the time series data $x_k(t)$ and $y_k(t)$, and the corresponding dual frequency RV-coupling coefficient given by Eq. 11. And consider the null hypothesis:

Eq. 27 $\quad H_0 : RV_{xy}(\theta,\beta) = 0$

The statistic for this test can be the actual estimated value for the dual frequency RV-coupling coefficient, denoted as:

Eq. 28 $\quad RV_{xy}(\theta,\beta)$

A simple non-parametric randomization procedure can be used to test $H_0$.

For instance, the method described by (Onslow et al., 2011) can be used, where basically the time sequence (i.e. the order in time) of one of the time series is randomly shuffled, and the statistic (Eq. 28) recomputed. The randomization is actually applied to time blocks, in such a way as to conserve the basic properties of the signal, while removing the relation (if any) to the other signal. The randomization is repeated many times, providing an estimator for the empirical probability distribution under the condition of no-association, which can then be used to estimate the probability of the statistic.

Furthermore, this procedure can be applied to the phase-phase and phase-amplitude-phase dual frequency RV-coupling coefficients defined above.

In the case when more than two frequency bands are considered, for instance θ, α, and β, then all possible pairs of cross-band couplings can be computed, and the maximum value of Eq. 28 among all band pairs should be used in the randomization procedure. The empirical probability distribution for the "max-statistic" gives a threshold with correction for multiple testing, see e.g. (Nichols and Holmes, 2002).

## 12. Toy example: cross-frequency phase-amplitude coupling

A simple toy example for phase-amplitude cross-frequency coupling between two signals is:

**Eq. 29** $\quad x(t) = \sin(\theta t + \phi_\theta) + \varepsilon_{xt}$

**Eq. 30** $\quad y(t) = [a + b\sin(\theta t + \phi_\theta)]\sin(\beta t + \phi_\beta) + \varepsilon_{yt}$

where θ and β are two frequencies, with $\theta < \beta$; $\phi_\theta$ and $\phi_\beta$ are phases, "a" and "b" are positive parameters satisfying $a \geq b$; and $\varepsilon_{xt}$ and $\varepsilon_{yt}$ are white noise. It is important to note that:
1. The "y" signal has the basic structure of a cyclostationary process.
2. This is a commonly used toy example in the literature for testing methods that quantify phase-amplitude coupling, see e.g. (Onslow et al., 2011; Berman et al., 2012; van Wijk et al., 2015).

The parameters used in the toy example here are:

**Eq. 31**
$$\begin{cases} \phi_\theta = \phi_\beta = 0 \\ a = 1.0 \\ b = 0.9 \\ \theta = 2 \times \pi \times 4/64 \\ \beta = 2 \times \pi \times 26/64 \\ \varepsilon_{xt}, \varepsilon_{yt} \; iid \sim N(\mu = 0, \sigma = 0.316) \\ t = 1...64000 \end{cases}$$

If it is assumed that the sampling rate is 64 Hz, then the theta frequency is 4Hz and the beta frequency is 26 Hz.

In summary, this simulation proceeded as follows:
1. Place two noisy source signals in the brain (signals from Eq. 29, Eq. 30, Eq. 31).
2. Generate EEG recordings at 19 electrodes.
3. Compute at 6239 cortical voxels, the current density signals, using eLORETA. This produces a trivariate time series at each voxel (three components of the current density vector field).

4. Use the two estimated trivariate time series at the locations of the two original source voxels and calculate the dual-frequency RV-coupling coefficient between all pairs of three frequency bands.

In detail, these signals (Eq. 29, Eq. 30, Eq. 31) were used as electric neuronal activity (current density) in a human head model, as follows:
- The "x" signal (theta) was assigned to a grey matter cortical voxel, located at (X= -25 , Y= 65 , Z= -5mm) (MNI coordinates); Brodmann area 10; Left Superior Frontal Gyrus
- The "y" signal (beta bursts occurring during the positive theta half cycles) was assigned to a grey matter cortical voxel, located at (X= 20 , Y= -100 , Z= 5mm) (MNI coordinates); Brodmann area 18; Left Middle Occipital Gyrus.

These two signals are shown in Figure 1.

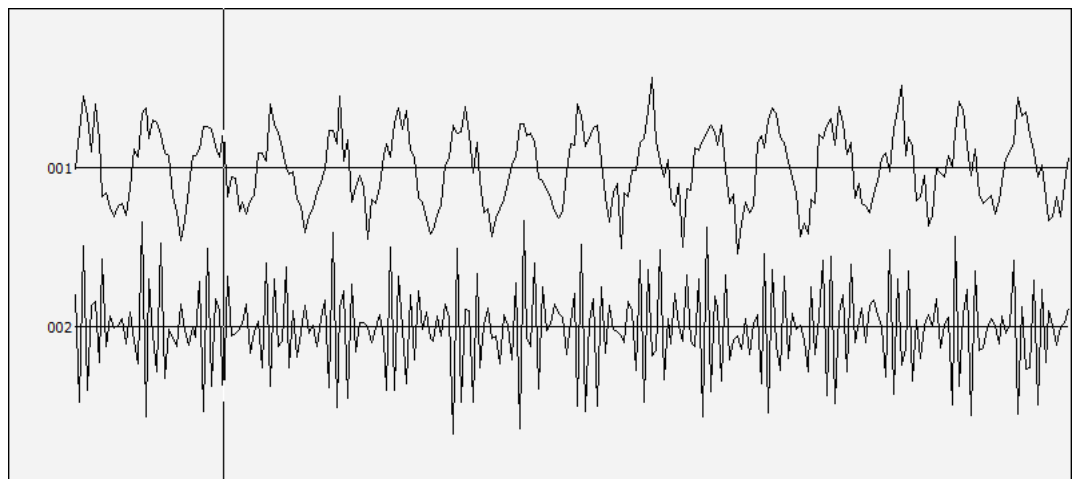

Figure 1: The two signals generated by Eq. 29 and Eq. 30, using the parameters in Eq. 31. The figure shows a total of 4 seconds (i.e. 256 time samples).

Using the forward EEG equations, extracranial EEG recordings were then computed at the 19 electrodes of the 10/20 system. Figure 2 show 4 seconds of EEG data.

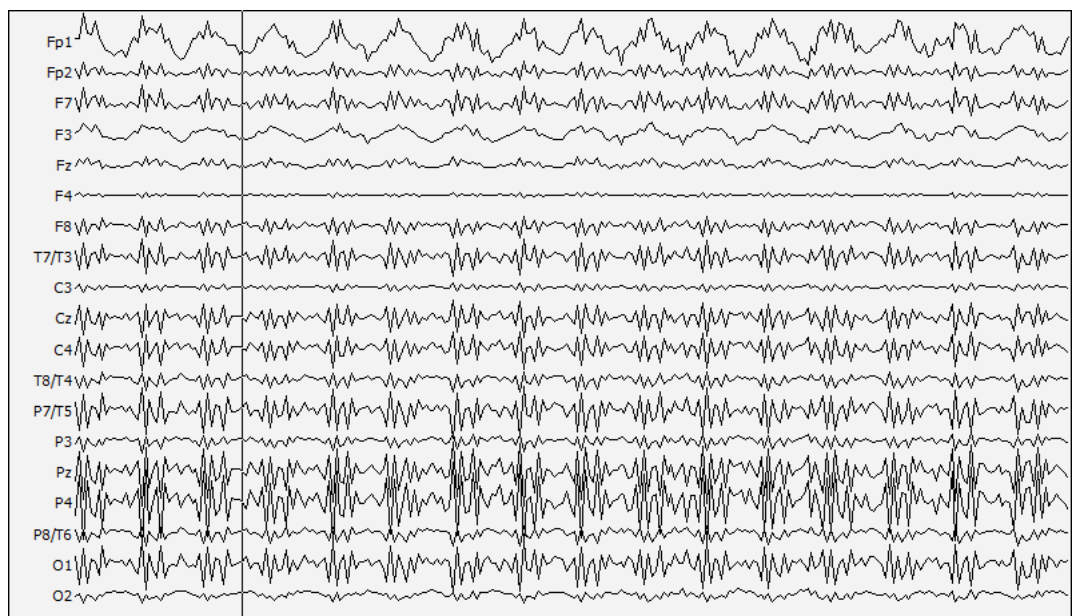

Figure 2: Four-seconds (256 time samples) of EEG generated from two sources, one in the left frontal cortex and the other in the right occipital cortex, with activity time series given by Eq. 29 and Eq. 30.

This toy example EEG data (19 time series with 64000 time samples) was then used for the estimation of the current density vector field all over the cortex (at 6239 voxels), using the eLORETA inverse solution (Pascual-Marqui et al., 2011). At each voxel, three time series are available, corresponding to the three components of the current density field.

Figure 3 and Figure 4 show scalp maps and eLORETA current density distributions at two moments in time, corresponding to positive and negative theta phase.

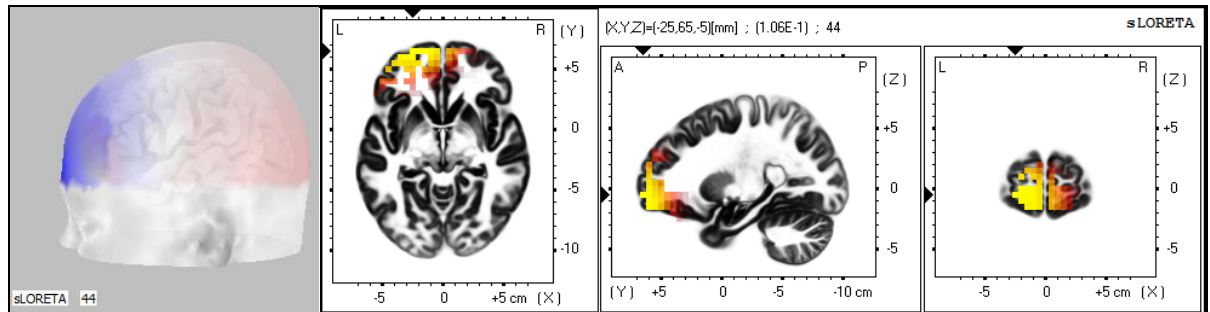

Figure 3: Scalp electric potential field (left panel, red positive, blue negative) and current density magnitude (slices in 3 right panels) at a time sample with negative left frontal theta phase and weak right posterior beta activity. At this time slice, the current density maximum is correctly located with eLORETA in left frontal cortex.

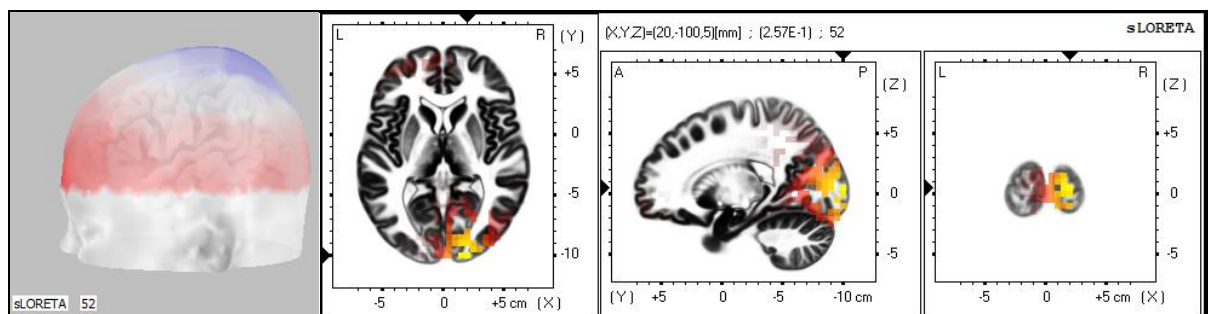

Figure 4: Scalp electric potential field (left panel, red positive, blue negative) and current density magnitude (slices in 3 right panels) at a time sample with positive left frontal theta phase and strong right posterior beta activity. At this time slice, the current density maximum is correctly located with eLORETA in right occipital cortex.

The estimated current density at the original two source voxels was then finally used for calculating the dual frequency RV-coupling coefficient between the pair of trivariate time series, for the following frequency bands:
Low: 1 to 8 Hz
Middle: 9 to 19 Hz
High: 20 to 30 Hz

The pair of estimated trivariate time series are shown in Figure 5.

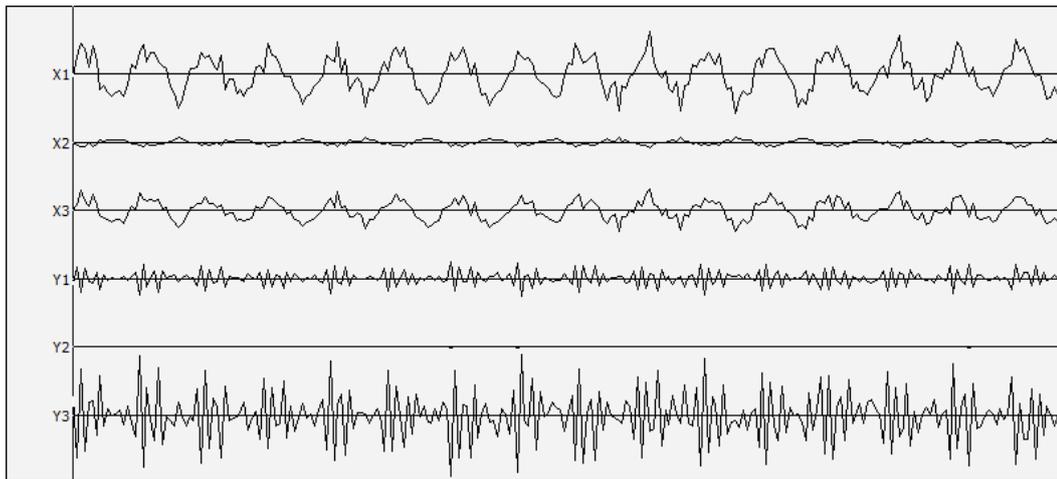

Figure 5: Intracranial signals of the current density vector field estimated with eLORETA. The two pairs (X and Y) of 3-dimensional signals correspond to left frontal and right occipital cortical locations.

Table 1 shows the estimated dual frequency RV-coupling coefficient values, between the trivariate frontal time series (theta) denoted as "**X**" and the trivariate occipital time series (beta burst only during positive theta phase) denoted as "**Y**".

|   |             | Y        |             |           |
|---|-------------|----------|-------------|-----------|
|   |             | Low freq | Middle freq | High freq |
| **X** | Low freq    |          | 0.0039      | 0.9892    |
|   | Middle freq | 0.0087   |             | 0.0039    |
|   | High freq   | 0.0122   | 0.0110      |           |

Table 1: Estimated dual frequency RV-coupling coefficient values, calculated between the trivariate frontal time series (theta) denoted as "**X**" and the trivariate occipital time series (beta burst only during positive theta phase) denoted as "**Y**".

As expected, from Table 1, the highlighted value for the dual frequency RV-coupling coefficient (0.9892) corresponds to frontal ("X") low frequency (which includes 4Hz theta) with posterior ("Y") high frequency (which includes 26 Hz beta). This is the only significant coupling, evaluated by means of the randomization procedure with correction for multiple testing. For 1000 randomizations, this produced p=0.001 for this coupling. All other couplings were not significant.

Note the following important result:
1. The data was generated by what is accepted in the literature as a case of "phase-amplitude coupling".
2. The simple dual frequency RV-coupling coefficient was used. This is not specific to phase-amplitude coupling. Instead, it is a general coupling measure that includes both amplitude and phase from both multivariate signals. Nevertheless, it detected correctly the coupling.